# Various angle periods of parabolic coincidence fringes in violation of Bell inequality with high-dimensional two-photon entanglement

Hsiao-Chih Huang[1,2,*]

[1]*Institute of Atomic and Molecular Sciences, Academia Sinica, Taipei 106, Taiwan*
[2]*HC Photonics Corporation, No. 2, Technology Rd. V, Hsinchu 300, Taiwan*
*Email: d93222016@ntu.edu.tw

**Abstract:** Two quantum states of two half-charge optical vortexes with relative azimuthal angle $\pi$ are orthogonal, by which two half-charge spiral phase plates with intersection angle $\pi$ can be used to demonstrate a parabolic coincident fringe in the Bell inequality experiment with high-dimensional two-photon orbital angular momentum entanglement and to thereby obtain a strong Bell parameter of $3\frac{1}{5}$. I theoretically demonstrate various orthogonal relations between two quantum states, each of which is a state with a rotational symmetry superposition made up of $n$ fractional orbital angular momentum states, where $n \in N$. I propose a Bell inequality experiment with two *n*-section spiral phase plates that have these two quantum states, whereby various angle periods with parabolic coincident fringes $2\pi/n$ and Bell's parameter $3\frac{1}{5}$ in each period can be obtained.

## I. INTRODUCTION

Over the past few decades, entangled photon pairs obtained by spontaneously parametric down-conversion (SPDC) [1, 2] have been used to experimentally verify the quantum nonlocality in the violation of the Bell inequality [3]. For two-particle and two-dimensional entanglement, the measured coincidence fringe of the pairs is the square of the sinusoid and Bell's parameter is $2\sqrt{2}$ [4], as proved using photon polarization [5], phase in spatial parity of one-dimensional transverse field (Hermite-Gaussian 10 mode) [6], phase with ghost imaging [7], and orbital angular momentum (OAM) of identical topological charge [8, 9]. In the first three cases, the angle periods of the coincidence fringes are respectively fixed at $\pi$, $2\pi$, rand $\pi$. However, in the last case, the angle periods vary over the range $\pi/m$, where $m$ is the topological charge. In this case, the period variable is attributed to the fact that the phase of the photon OAM can vary by $m$ multiples of $2\pi$ in an azimuthal rotation cycle [10, 11], whereby $m$ pairs of analyzers that have the property of $m$-fold rotational symmetry are used [8, 9]. By contrast, the single-value period $\pi$ in the first case is attributed to the fact that the orientation of photon polarization only varies by $\pi$ in a rotation cycle, whereby two polarizers, a pair of analyzers of two-fold rotational symmetry, are used. The reasons for those single-value periods in the second and third cases are similar, whereby the respectively vital pair of analyzers are used. A variety of periods should be significant because a large period results in high angle resolution of the coincident fringe, whereas a small period results in a small range requirement of the relative orientation between the two analyzers.

Nevertheless, the OAM entanglement of twin photons obtained by SPDC can have infinite dimensions [12, 13]. The higher dimension indicates that the violation is more robust against noise [14]. Dada *et al.* demonstrated violations of the generalized Bell inequality [15] with two-photon and up to 12-arbitrary-dimensional entanglement by using a setup with two spatial light modulations [16] that are configured by the hologram generation algorithm [17]. Oemrawsingh *et al.* demonstrated a violation of the Clauser-Horne-Shimony-Holt (CHSH) Bell inequality with two-photon and high-dimensional entanglement through two detections with two half-charge spiral phase plates (SPPs), and they obtained the outcomes of the coincidence fringe of the parabola and a strong Bell's parameter of $3\frac{1}{5}$ [18, 19]. One key factor in verifying the quantum nonlocality with a parabolic coincidence fringe is that the quantum state of a light beam with fractional OAM (fractional optical vortex) is the superposition state consisting of numerous optical OAM eigenmodes [20] with a functional weight [21]. However, the periods of both of these coincident fringes are only $2\pi$. In the latter case, the monotonic period is attributable to the rotational asymmetry of the SPP structure.

In this article, I theoretically demonstrate the overlap probability between two quantum states, each of which is a state with a rotational symmetry superposition made up of $n$ fractional OAM states, is $n$-section parabolas in a cycle round $2\pi$. Various orthogonal relations exist between these two quantum states with charges of $\left(M - \frac{n}{2}\right) \bmod n = 0$ and rotation angles of $\alpha = \pi(2t-1)/n$, $t = 1, 2, ..., n$. The superposed quantum state has the property of *n*-fold rotational symmetry and is still the superposition state consisting of numerous optical OAM eigenmodes with functional weights [14]. I propose an experiment to prove the violations of the Bell inequality with high-dimensional two-particle OAM entanglement using an SPDC setup with two analyzers that have these two quantum states; however, the periods of the parabolic coincident fringes vary as $2\pi/n$, and Bell's parameter of $3\frac{1}{5}$ is obtained in each period. Thus, the requirement of small angle range of $2\pi/n$ can be used to prove the quantum nonlocality with high-dimensional two-photon entanglement. These two analyzers consist of two *n*-section SPPs, and the period variable of parabolic coincident fringes is attributable to the fact that the structure of an *n*-section SPP has the property of *n*-fold rotational symmetry.

## II. THEORY

An SPP with *n* edge dislocations, denoted herein as SPP*n*, is an *n*-section SPP [22]. The SPP*n* structure has the property of *n*-fold rotational symmetry. Figure 1 shows the two cases of *n* = 2 and 4, that is, SPP2 and SPP4, respectively. These are two- and four-section transparent dielectric plates, respectively, whose thicknesses vary linearly with the azimuthal angle $\phi$, and therefore, they shift the phases of the incident light fields in two- and four-fold symmetries, respectively. Without loss of generality, two indexes $M$ and $\alpha$ are used to respectively represent the thickness gradient with respect to $\phi$ and the orientation of the edge dislocation of an SPP*n* [18, 21]. Owing to the angle repetition for a rotation of $2\pi$, $\alpha$ is bounded by $0 \leq \alpha < 2\pi$. An SPP*n* can serve as an operator similar to the SPP in the basis set transformation for optical OAM eigenmodes in the Bell inequality experiment [18]. By operating an SPP*n* operator $\hat{S}_n(\alpha, M)$ on the integer OAM basis $\{|m'\rangle\}$ [23], a new basis $\{|Mn^{(m')}(\alpha)\rangle\}$ is expressed as $\hat{S}_n(\alpha, M)|m'\rangle \equiv |Mn^{(m')}(\alpha)\rangle$. The azimuthal part of a light field with OAM eigenvalue $m'$ after this operation is $\langle\phi|\hat{S}_n(\alpha, M)|m'\rangle = e^{im'\phi}\langle\phi|Mn(\alpha)\rangle$. This is equal to one phase term $e^{im'\phi}$ times the other one $\langle\phi|Mn(\alpha)\rangle$ that has the phase shift structure made from an SPP*n*. It is readily seen that $|Mn(\alpha)\rangle$ is the quantum state of a light beam that is the exportation by passing a fundamental Gaussian light beam through an SPP*n*. For $n = 1$, by using an SPP, this is the quantum state of a fractional optical vortex with fractional topological charge $M$, $|M(\alpha)\rangle$ [21].

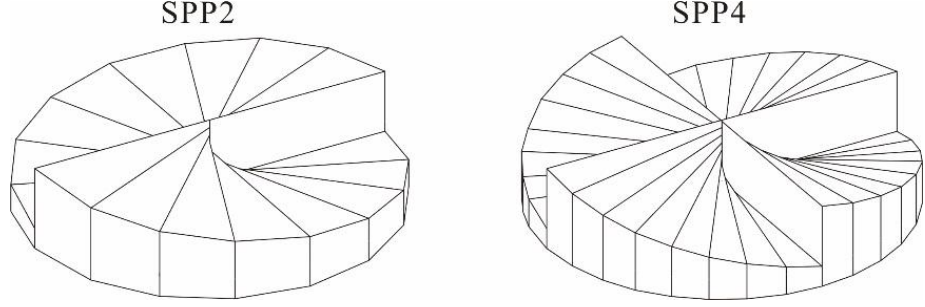

Fig. 1. Schemas of SPP2 and SPP4.

The phase gradients of the helical wavefront with respect to $\phi$ are invariant for azimuthally superposing fractional optical vortexes with identical charges [14] as well as that with integer charge $m$ [24]. The quantum state follows a similar principle; in having identical $M$, $|Mn(\alpha)\rangle$ with $M$ must be given with the rotational symmetry superposition made up of $n$ fractional OAM states $|M(\alpha)\rangle$ with $M$ by

$$|Mn\rangle = \frac{|Mn'\rangle}{\sqrt{\langle Mn'|Mn'\rangle}}, \quad |Mn'\rangle = \sum_{k=0}^{n-1} \hat{U}\left(2\pi \times \frac{k}{n}\right)|M\rangle, \quad (1)$$

where $\hat{U}$ is the rotation operator, as introduced in Appendix A, and $|Mn'\rangle$ is the unnormalized quantum state, whose field amplitudes with superposition of $n = 2$ and 4 are derived in Appendix B. Along with the structure of SPP$n$, $|Mn(\alpha)\rangle$ has the property of $n$-fold rotational symmetry, which is proven theoretically in Appendix C, and the rotational symmetry of the structured light beams with $|M2\rangle$ and $|M4\rangle$ has been verified experimentally in terms of intensity image and phase profile in Ref. 14. $|Mn(\alpha)\rangle$, similar to $|M(\alpha)\rangle$ [20, 21], is a superposition state consisting of numerous optical OAM eigenmodes with functional weights [24], although $|Mn(\alpha)\rangle$ has vanished OAM components as $m'$ mod $n$ [14]. A large $n$ implies a large rotational symmetry for both the structure of SPP$n$ and $|Mn(\alpha)\rangle$. Let $\alpha = 0$ and $\{|Mn^{(m')}(0)\rangle\}$ be complete because $\hat{S}_n(\alpha,M)$ is a unitary operator (an invariant field state that is passing through two SPP$n$s with identical $M$ and $n$ but handedness reversal). $|Mn^{(m')}(\alpha)\rangle$ can be a superposition of these basis states, and $\alpha$ is equal to the intersection angle between two SPP$n$s. Thus, the decomposition of $|Mn^{(m')}(\alpha)\rangle$ into $\{|Mn^{(m')}(0)\rangle\}$ depends on $\alpha$. The sum of these projection scalars is equal to the overlap amplitude $\langle Mn(\alpha)|Mn(0)\rangle$, whose modulus square, the overlap probability, is $|\langle Mn(0)|Mn(\alpha)\rangle|^2$. The formulas derived in Appendix D are used to calculate the overlap probability with $n$.

For $n = 1$, the overlap probability between the quantum state of a light beam with fractional OAM and its rotation state with $\alpha$ is [18]

$$|\langle M(0)|M(\alpha)\rangle|^2 = \left(1-\frac{\alpha}{\pi}\right)^2 \sin^2(M\pi) + \cos^2(M\pi), \quad 0 \leq \alpha < 2\pi, \quad (2)$$

where Eq. (A3) is used.

The overlap amplitude for $n = 2$ is

$$\langle M2(0)|M2(\alpha)\rangle = \frac{\langle M(0)|M(\alpha)\rangle + \langle M(0)|\hat{U}(\pi)|M(\alpha)\rangle}{1+\cos(\pi M)}, \quad (3)$$

where Eqs. (B1) and (D3) are used. The $n = 2$ overlap probability is equal to the modulus square of the right hand side of Eq. (3) as

$$|\langle M2(0)|M2(\alpha)\rangle|^2 = \{|\langle M(0)|M(\alpha)\rangle|^2 + |\langle M(0)|\hat{U}(\pi)|M(\alpha)\rangle|^2 \\ + 2\operatorname{Re}[\langle M(0)|M(\alpha)\rangle\langle M(0)|\hat{U}(\pi)|M(\alpha)\rangle^*]\}/[1+\cos(\pi M)]^2. \quad (4)$$

Let $\beta = \pi$ in Eqs. (D7) and (D5), which are respectively expressed as

$$|\langle M(0)|\hat{U}(\pi)|M(\alpha)\rangle|^2 = \frac{1}{\pi^2} \times \begin{cases} (\alpha^2 - \pi^2)\sin^2(M\pi) + \pi^2, & 0 \leq \alpha < \pi \\ (\alpha-\pi)(\alpha-3\pi)\sin^2(M\pi) + \pi^2, & \pi \leq \alpha < 2\pi \end{cases} \quad (5)$$

and

$$\operatorname{Re}[\langle M(0)|M(\alpha)\rangle\langle M(0)|\hat{U}(\pi)|M(\alpha)\rangle^*] \\ = \frac{1}{4\pi^2} \times \begin{cases} \alpha(\pi-\alpha)\cos(3M\pi) + (4\pi^2 - \pi\alpha + \alpha^2)\cos(M\pi), & 0 \leq \alpha < \pi \\ (2\pi-\alpha)(\alpha-\pi)\cos(3M\pi) + (6\pi^2 - 3\pi\alpha + \alpha^2)\cos(M\pi), & \pi \leq \alpha < 2\pi \end{cases}. \quad (6)$$

Substituting Eqs. (2), (5), and (6) into Eq. (4) gives

$$|\langle M2(0)|M2(\alpha)\rangle|^2 = \begin{cases} \left(1-\frac{2\alpha}{\pi}\right)^2 \sin^2\left(\frac{M\pi}{2}\right)+\cos^2\left(\frac{M\pi}{2}\right), & 0\leq\alpha<\pi \\ \left(3-\frac{2\alpha}{\pi}\right)^2 \sin^2\left(\frac{M\pi}{2}\right)+\cos^2\left(\frac{M\pi}{2}\right), & \pi\leq\alpha<2\pi \end{cases}. \quad (7)$$

The overlap amplitude for $n = 4$ is

$$\langle M4(0)|M4(\alpha)\rangle = \left[\langle M(0)|M(\alpha)\rangle + \langle M(0)|\hat{U}\left(\tfrac{\pi}{2}\right)|M(\alpha)\rangle + \langle M(0)|\hat{U}(\pi)|M(\alpha)\rangle \right. \\ \left. + \langle M(0)|\hat{U}\left(\tfrac{3\pi}{2}\right)|M(\alpha)\rangle\right] / \left[1+\cos(\pi M)+2\cos^3\left(\tfrac{\pi}{2}M\right)\right], \quad (8)$$

where Eqs. (B2) and (D3) are used. The $n = 4$ overlap probability is equal to the modulus square of the right hand side of Eq. (8) as

$$|\langle M4(0)|M4(\alpha)\rangle|^2 = \Big\{ |\langle M(0)|M(\alpha)\rangle|^2 + |\langle M(0)|\hat{U}(\pi)|M(\alpha)\rangle|^2 + |\langle M(0)|\hat{U}\left(\tfrac{\pi}{2}\right)|M(\alpha)\rangle|^2 \\ + |\langle M(0)|\hat{U}\left(\tfrac{3\pi}{2}\right)|M(\alpha)\rangle|^2 + 2\mathrm{Re}\left[\langle M(0)|M(\alpha)\rangle\langle M(0)|\hat{U}(\pi)|M(\alpha)\rangle^*\right] \\ + 2\mathrm{Re}\left[\langle M(0)|M(\alpha)\rangle\langle M(0)|\hat{U}\left(\tfrac{\pi}{2}\right)|M(\alpha)\rangle^*\right] + 2\mathrm{Re}\left[\langle M(0)|M(\alpha)\rangle\langle M(0)|\hat{U}\left(\tfrac{3\pi}{2}\right)|M(\alpha)\rangle^*\right] \\ + 2\mathrm{Re}\left[\langle M(0)|\hat{U}(\pi)|M(\alpha)\rangle\langle M(0)|\hat{U}\left(\tfrac{\pi}{2}\right)|M(\alpha)\rangle^*\right] + 2\mathrm{Re}\left[\langle M(0)|\hat{U}(\pi)|M(\alpha)\rangle\langle M(0)|\hat{U}\left(\tfrac{3\pi}{2}\right)|M(\alpha)\rangle^*\right] \\ + 2\mathrm{Re}\left[\langle M(0)|\hat{U}\left(\tfrac{\pi}{2}\right)|M(\alpha)\rangle\langle M(0)|\hat{U}\left(\tfrac{3\pi}{2}\right)|M(\alpha)\rangle^*\right] \Big\} / \left[1+\cos(\pi M)+2\cos^3\left(\tfrac{\pi}{2}M\right)\right]^2.$$

(9)

Substituting $\beta = \pi/2$ and $3\pi/2$ in Eq. (D7) gives

$$|\langle M(0)|\hat{U}\left(\tfrac{\pi}{2}\right)|M(\alpha)\rangle|^2 = \frac{1}{\pi^2} \times \begin{cases} \left(\alpha+\tfrac{\pi}{2}\right)\left(\alpha-\tfrac{3\pi}{2}\right)\sin^2(M\pi)+\pi^2, & 0\leq\alpha<\tfrac{3\pi}{2} \\ \left(\alpha-\tfrac{3\pi}{2}\right)\left(\alpha-\tfrac{7\pi}{2}\right)\sin^2(M\pi)+\pi^2, & \tfrac{3\pi}{2}\leq\alpha<2\pi \end{cases} \quad (10)$$

and

$$|\langle M(0)|\hat{U}\left(\tfrac{3\pi}{2}\right)|M(\alpha)\rangle|^2 = \frac{1}{\pi^2} \times \begin{cases} \left(\alpha+\tfrac{3\pi}{2}\right)\left(\alpha-\tfrac{\pi}{2}\right)\sin^2(M\pi)+\pi^2, & 0\leq\alpha<\tfrac{\pi}{2} \\ \left(\alpha-\tfrac{\pi}{2}\right)\left(\alpha-\tfrac{5\pi}{2}\right)\sin^2(M\pi)+\pi^2, & \tfrac{\pi}{2}\leq\alpha<2\pi \end{cases}. \quad (11)$$

Substituting $\beta = \pi/2$ and $3\pi/2$ in Eq. (D5) gives

$$\mathrm{Re}\left[\langle M(0)|M(\alpha)\rangle\langle M(0)|\hat{U}\left(\tfrac{\pi}{2}\right)|M(\alpha)\rangle^*\right] \\ = \frac{1}{4\pi^2} \times \begin{cases} \left[2\left(2\alpha^2-3\pi\alpha-\pi^2\right)\sin^2(M\pi)+4\pi^2\right]\cos\left(\tfrac{M\pi}{2}\right)+\pi^2\sin(2M\pi)\sin\left(\tfrac{M\pi}{2}\right), & 0\leq\alpha<\tfrac{3\pi}{2} \\ \left(2\alpha^2-7\pi\alpha+7\pi^2\right)\cos\left(\tfrac{3M\pi}{2}\right)+\left(\tfrac{7}{2}\pi\alpha-\alpha^2-3\pi^2\right)\cos\left(\tfrac{7M\pi}{2}\right)+\alpha\left(\tfrac{7}{2}\pi-\alpha\right)\cos\left(\tfrac{M\pi}{2}\right), & \tfrac{3\pi}{2}\leq\alpha<2\pi \end{cases}$$

(12)

and

$$\operatorname{Re}\left[\langle M(0)|M(\alpha)\rangle\langle M(0)|\hat{U}\left(\tfrac{3\pi}{2}\right)|M(\alpha)\rangle^{*}\right]$$

$$=\frac{1}{4\pi^{2}}\times\begin{cases}\left[2\left(2\alpha^{2}-\pi\alpha+3\pi^{2}\right)\sin^{2}(M\pi)+4\pi^{2}\right]\cos\left(\tfrac{3M\pi}{2}\right)+3\pi^{2}\sin(2M\pi)\sin\left(\tfrac{3M\pi}{2}\right), & 0\le\alpha<\tfrac{\pi}{2}\\ \left(2\alpha^{2}-5\pi\alpha+5\pi^{2}\right)\cos\left(\tfrac{M\pi}{2}\right)+\left(\tfrac{5}{2}\pi\alpha-\alpha^{2}-\pi^{2}\right)\cos\left(\tfrac{5M\pi}{2}\right)+\alpha\left(\tfrac{5}{2}\pi-\alpha\right)\cos\left(\tfrac{3M\pi}{2}\right), & \tfrac{\pi}{2}\le\alpha<2\pi\end{cases}.$$

(13)

Substituting $\beta_{1}=\pi$ and $\beta_{2}=\pi/2$, $\beta_{1}=\pi$ and $\beta_{2}=3\pi/2$, and $\beta_{1}=\pi/2$ and $\beta_{2}=3\pi/2$ in Eq. (D8) gives

$$\operatorname{Re}\left[\langle M(0)|\hat{U}(\pi)|M(\alpha)\rangle\langle M(0)|\hat{U}\left(\tfrac{\pi}{2}\right)|M(\alpha)\rangle^{*}\right]=$$

$$\frac{1}{4\pi^{2}}\times\begin{cases}\left[4\sin^{2}(M\pi)\left(\alpha^{2}-\tfrac{\pi}{2}\alpha-\pi^{2}\right)+4\pi^{2}\right]\cos\left(\tfrac{M\pi}{2}\right)+\pi^{2}\sin(2M\pi)\sin\left(\tfrac{M\pi}{2}\right), & 0\le\alpha<\pi\\ \left(2\alpha^{2}+4\pi^{2}-5\pi\alpha\right)\cos\left(\tfrac{3M\pi}{2}\right)+(\alpha-\pi)\left(\tfrac{3\pi}{2}-\alpha\right)\cos\left(\tfrac{7M\pi}{2}\right)+(3\pi-\alpha)\left(\alpha+\tfrac{\pi}{2}\right)\cos\left(\tfrac{M\pi}{2}\right), & \pi\le\alpha<\tfrac{3\pi}{2},\\ \left[4\sin^{2}(M\pi)\left(\alpha^{2}-\tfrac{9}{2}\pi\alpha+4\pi^{2}\right)+4\pi^{2}\right]\cos\left(\tfrac{M\pi}{2}\right)+\pi^{2}\sin(2M\pi)\sin\left(\tfrac{M\pi}{2}\right), & \tfrac{3\pi}{2}\le\alpha<2\pi\end{cases}$$

(14)

$$\operatorname{Re}\left[\langle M(0)|\hat{U}(\pi)|M(\alpha)\rangle\langle M(0)|\hat{U}\left(\tfrac{3}{2}\pi\right)|M(\alpha)\rangle^{*}\right]=$$

$$\frac{1}{4\pi^{2}}\times\begin{cases}\left[4\sin^{2}(M\pi)\left(\alpha^{2}+\tfrac{\pi}{2}\alpha-\pi^{2}\right)+4\pi^{2}\right]\cos\left(\tfrac{M\pi}{2}\right)+\pi^{2}\sin(2M\pi)\sin\left(\tfrac{M\pi}{2}\right), & 0\le\alpha<\tfrac{\pi}{2}\\ \left(2\alpha^{2}+2\pi^{2}-3\pi\alpha\right)\cos\left(\tfrac{3M\pi}{2}\right)+(\pi-\alpha)\left(\alpha-\tfrac{\pi}{2}\right)\cos\left(\tfrac{7M\pi}{2}\right)+(\alpha+\pi)\left(\tfrac{5\pi}{2}-\alpha\right)\cos\left(\tfrac{M\pi}{2}\right), & \tfrac{\pi}{2}\le\alpha<\pi\\ \left[4\sin^{2}(M\pi)\left(\alpha^{2}-\tfrac{7\pi}{2}\alpha+2\pi^{2}\right)+4\pi^{2}\right]\cos\left(\tfrac{M\pi}{2}\right)+\pi^{2}\sin(2M\pi)\sin\left(\tfrac{M\pi}{2}\right), & \pi\le\alpha<2\pi\end{cases}$$

(15)

, and

$$\operatorname{Re}\left[\langle M(0)|\hat{U}\left(\tfrac{\pi}{2}\right)|M(\alpha)\rangle\langle M(0)|\hat{U}\left(\tfrac{3\pi}{2}\right)|M(\alpha)\rangle^{*}\right]=$$

$$\frac{1}{4\pi^{2}}\times\begin{cases}\left[4\sin^{2}(M\pi)\left(\alpha^{2}-\tfrac{5}{4}\pi^{2}\right)+4\pi^{2}\right]\cos(M\pi)+2\pi^{2}\sin(2M\pi)\sin(M\pi), & 0\le\alpha<\tfrac{\pi}{2}\\ \left(\alpha^{2}-2\pi\alpha+\tfrac{19}{4}\pi^{2}\right)\cos(M\pi)+\left(\tfrac{3\pi}{2}-\alpha\right)\left(\alpha-\tfrac{\pi}{2}\right)\cos(3M\pi), & \tfrac{\pi}{2}\le\alpha<\tfrac{3\pi}{2}\\ \left[4\sin^{2}(M\pi)\left(\alpha^{2}-4\pi\alpha+\tfrac{11}{4}\pi^{2}\right)+4\pi^{2}\right]\cos(M\pi)+2\pi^{2}\sin(2M\pi)\sin(M\pi), & \tfrac{3\pi}{2}\le\alpha<2\pi\end{cases}.$$

(16)

Substituting Eqs. (2), (5), (6), and (10)–(16) into Eq. (9) gives

$$|\langle M4(0)|M4(\alpha)\rangle|^2 = \begin{cases} \left(1-\frac{4\alpha}{\pi}\right)^2 \sin^2\left(\frac{M\pi}{4}\right) + \cos^2\left(\frac{M\pi}{4}\right), & 0 \leq \alpha < \frac{\pi}{2} \\ \left(3-\frac{4\alpha}{\pi}\right)^2 \sin^2\left(\frac{M\pi}{4}\right) + \cos^2\left(\frac{M\pi}{4}\right), & \frac{\pi}{2} \leq \alpha < \pi \\ \left(5-\frac{4\alpha}{\pi}\right)^2 \sin^2\left(\frac{M\pi}{4}\right) + \cos^2\left(\frac{M\pi}{4}\right), & \pi \leq \alpha < \frac{3\pi}{2} \\ \left(7-\frac{4\alpha}{\pi}\right)^2 \sin^2\left(\frac{M\pi}{4}\right) + \cos^2\left(\frac{M\pi}{4}\right), & \frac{3\pi}{2} \leq \alpha < 2\pi \end{cases}. \quad (17)$$

According to Eqs. (2), (7), and (17), the overlap probability for arbitrary $n$ should be calculated as

$$|\langle Mn(0)|Mn(\alpha)\rangle|^2 = \frac{[\pi(2t-1)-n\alpha]^2}{\pi^2} \sin^2\left(\frac{M\pi}{n}\right) + \cos^2\left(\frac{M\pi}{n}\right),$$
$$\frac{2\pi}{n}(t-1) \leq \alpha < \frac{2\pi}{n}t,\ t = 1, 2, ..., n. \quad (18)$$

This probability generally depends on the integer part of the step index $m$; however, it is independent of the topological charge of twin photons $m'$. In particular, it is independent of $m$ and $m'$ as $n = 1$ [18]. Eq. (18) is a constant function of $\alpha$ as $M \bmod n = 0$, and it is otherwise a parabolic function of $\alpha$ that is repeated between $n$ sections as $M \bmod n \neq 0$. These $n$-section parabolic functions, similar to previously reported parabolic functions [18], indicates that SPP$n$ rotates the OAM state following through the Hilbert space that is not confined to a two-dimensional subspace. There are orthogonal relations such as $\left(M-\frac{n}{2}\right) \bmod n = 0$ at $\alpha = \pi(2t-1)/n$. Equivalently, all the total phase variations in each fold of these superposed quantum states with various $n$ cases are module $2\pi$ by $\pi$ in these orthogonal relations. Two probability functions with $n = 2$ and $n = 4$ are plotted in Fig. 2 for cases of uncorrelated, correlated, and orthogonal relations.

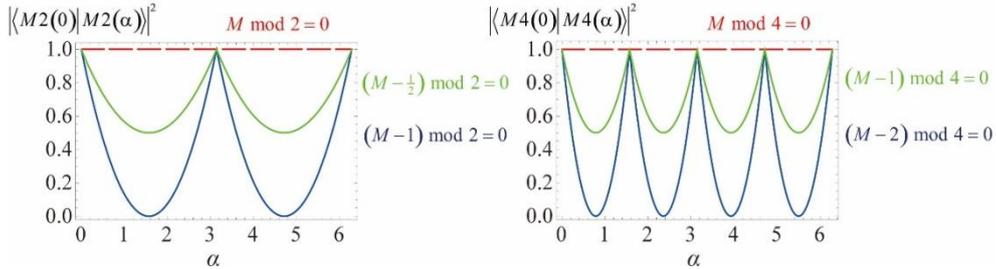

Fig. 2. Various angle periods of overlap probability. Left: three overlap probabilities between a superposed quantum state $|M2(0)\rangle$ and its rotation $|M2(\alpha)\rangle$ in the charges $M \bmod 2 = 0$, $\left(M-\frac{1}{2}\right) \bmod 2 = 0$, and $(M-1) \bmod 2 = 0$, as indicated respectively by red dotted, green, and blue solid curves. The first is a constant, whereas the latter two are parabolic functions of

$\alpha$ down to 0.5 and 0 at $\alpha = \pi/2$ and $3\pi/2$. Right: three overlap probabilities between a superposed quantum state $|M4(0)\rangle$ and its rotation $|M4(\alpha)\rangle$ in the charges $M \mod 4 = 0$, $(M-1) \mod 4 = 0$, and $(M-2) \mod 4 = 0$. The first is a constant, whereas the latter two are parabolic functions of $\alpha$ down to 0.5 and 0 at $\alpha = \pi/4, 3\pi/4, 5\pi/4$, and $7\pi/4$.

### III. PROPOSED EXPERIMENT

I propose an experimental setup similar to the one in Figure 2 of Ref. 18 but with the two SPPs with half-integers being replaced by two SPP$n$s with $(M - \frac{n}{2}) \mod n = 0$ to prove the violations of the Bell inequality with high-dimensional two-particle OAM entanglement. This is a type of SPDC setup. Its generated twin photons in two arms, signal and idler, obey OAM conservation [12] provided that two conditions of the paraxial limit and thin-crystal approximation [25-28]. By choosing a fundamental Gaussian pump beam in an SPDC setup with photon OAM conservation, the normalized coincidence fringe detected by the two photons after passing two SPP$n$s that are intersected by $\alpha$ is equal to the mode overlap probability between two quantum states (Eq. (18)), as also described in Refs. 18 and 19. However, the fringes are $n$-section parabolas. These expected outcomes with various azimuthal angle periods of parabolic coincident fringes will have a small range of intersection angles in proving quantum nonlocality with high-dimensional two-photon entanglement.

As in the case of two SPPs [18], the CHSH inequality is chosen in my proposed experiment as $S = E(\alpha_s, \alpha_i) - E(\alpha'_s, \alpha_i) + E(\alpha_s, \alpha'_i) + E(\alpha'_s, \alpha'_i) \leq 2$ [4], where $s$ and $i$ denote the signal and idler in the SPDC experiment, respectively. $E$ is expressed as [5, 29]

$$E(\alpha_s, \alpha_i) = \frac{P(\alpha_s, \alpha_i) + P(\alpha_s^\perp, \alpha_i^\perp) - P(\alpha_s, \alpha_i^\perp) - P(\alpha_s^\perp, \alpha_i)}{P(\alpha_s, \alpha_i) + P(\alpha_s^\perp, \alpha_i^\perp) + P(\alpha_s, \alpha_i^\perp) + P(\alpha_s^\perp, \alpha_i)}, \quad (19)$$

where $P(\alpha_s, \alpha_i)$ is the coincidence probability function proportional to $\left[1 - |\alpha_s - \alpha_i|/(\pi/n)\right]^2$ in the orthogonal case of $(M - \frac{n}{2}) \mod n = 0$ and only the modulus of $\alpha_s - \alpha_i$. $\alpha_s^\perp$ and $\alpha_i^\perp$ are the angles of two analyzers for the states orthogonal to the states with setting $\alpha_s$ and $\alpha_i$, respectively. In this case, there are $n$ pairs of them as $\alpha_s^\perp \equiv \alpha_s + \pi(2t-1)/n$ and $\alpha_i^\perp \equiv \alpha_i + \pi(2t-1)/n$, $t = 1, 2, ..., n$. As the periodicity in this case is $1/2n$ that in the case of photon polarization entanglement, I use the standard analyzer settings of a previous case [5, 30] divided by $2n$: $\alpha_s = -\pi/4n$, $\alpha'_s = \pi/4n$, $\alpha_i = -\pi/2n$, and $\alpha'_i = 0$. Upon substituting these parameters into Eq. (19), Bell's parameter in each of the $n$-sections can be evaluated as $3\frac{1}{5}$, which is identical to that mentioned in Ref. 18.

### IV. CONCLUSION

I have theoretically demonstrated the overlap probability between two quantum states, each of which is a state with a rotational symmetry superposition made up of $n$ fractional OAM states, is $n$-section parabolas. There are various orthogonal relations between these two quantum states at charges of $(M-\frac{n}{2}) \bmod n = 0$ and relative rotation angle $\alpha = \pi(2t-1)/n$, $t = 1, 2, ..., n$. I propose to prove the violations of the Bell inequality with high-dimensional two-particle OAM entanglement in an SPDC experiment with two SPP$n$s where the intersection angle represents various periods of parabolic coincident fringes. The occurring charge of $(M-\frac{n}{2}) \bmod n = 0$ is coincident with the occurrence of the completely destructive interference [24] (cf. Ref. 24, except $M$ = 1/2) and maximum phase dislocation degree for a light beam with phase singularity that is characterized with azimuthally symmetric angular positions [14].

**APPENDIX A: QUANTUM STATE OF A LIGHT BEAM WITH FRACTIONAL OAM**

The quantum state of a light beam with fractional OAM is denoted by $|M(\alpha)\rangle$ [21], where $M = m + \mu$ and the parameter $\alpha$, bounded by $0 \le \alpha < 2\pi$, is the angular position of the discontinuity. A function $f_\alpha(\phi)$ is introduced as

$$f_\alpha(\phi) = \begin{cases} 1, & 0 \le \phi < \alpha \\ 0, & \alpha \le \phi < 2\pi \end{cases}. \tag{A1}$$

By using Eq. (A1), a definition can be shown as

$$\langle \phi | M(\alpha) \rangle \equiv e^{im\phi} e^{i\mu[\phi + 2\pi f_\alpha(\phi) - \alpha]}. \tag{A2}$$

Based on the completeness relation and Eq. (A2), the overlap amplitude between two states with an intersection angle $\alpha$ is

$$\begin{aligned} \langle M(0) | M(\alpha) \rangle &= \frac{1}{2\pi} \int_0^{2\pi} d\phi\, e^{i2\pi\mu[f_\alpha(\phi) - f_0(\phi)]} e^{i\mu(0-\alpha)} \\ &= \frac{1}{2\pi} \left[ \alpha e^{i(2\pi-\alpha)\mu} + (2\pi - \alpha) e^{-i\alpha\mu} \right]. \end{aligned} \tag{A3}$$

$|M(\alpha)\rangle$ can be decomposed into integer OAM eigenmodes, the probability distribution of which can be obtained by setting $\langle M(0)| = \langle m'|$ in Eq. (A3)

$$c_{m'}[M(\alpha)] = \langle m' | M(\alpha) \rangle = \frac{i e^{i(m-m')\alpha}}{2\pi(M-m')} \left(1 - e^{i2\pi\mu}\right). \tag{A4}$$

The state resulting from a unitary operator $\hat{U}(\beta)$ is the effect of the rotation for this quantum state and an additional phase term $e^{-im\beta}$:

$$\hat{U}(\beta) | M(\alpha) \rangle = e^{-im\beta} | M(\alpha \oplus \beta) \rangle, \tag{A5}$$

where the parameter $\beta$ bounded by $0 \le \beta < 2\pi$ is the action on the state and $\alpha \oplus \beta = (\alpha + \beta) \bmod 2\pi$ yields a result in the range $[0, 2\pi)$ owing to the $2\pi$ modulo. The multiplication of rotation operators has the combination characteristic

$$\hat{U}(\beta_1) \hat{U}(\beta_2) | M(\alpha) \rangle = \hat{U}(\beta_1 \oplus \beta_2) | M(\alpha) \rangle. \tag{A6}$$

Some useful formulas are derived as follows.

$$\langle M(\alpha)|\hat{U}(\beta)|M(\alpha)\rangle = \frac{e^{-im\beta}}{2\pi}\left[\beta e^{i(2\pi-\beta)\mu}+(2\pi-\beta)e^{-i\beta\mu}\right], \quad (A7)$$

where Eqs. (A3) and (A5) are used. The real part of Eq. (A7) is

$$\text{Re}\langle M(\alpha)|\hat{U}(\beta)|M(\alpha)\rangle = \frac{1}{2\pi}\left[\beta\cos(2\pi\mu-\beta M)+(2\pi-\beta)\cos(\beta M)\right]. \quad (A8)$$

By using Eqs. (A4) and (A5),

$$\langle m'|\hat{U}(\beta)|M(\alpha)\rangle = e^{-im\beta}\frac{ie^{i(m-m')(\beta\oplus\alpha)}}{2\pi(M-m')}\left(1-e^{i2\pi\mu}\right). \quad (A9)$$

## APPENDIX B: USEFUL FORMULAS FOR QUANTUM STATES WITH SUPERPOSITIONS OF $n = 2$ AND $4$

In Eq. (1), the field amplitude of the unnormalized quantum state with $n = 2$ superposition is

$$\begin{aligned}\langle M2'|M2'\rangle &= \left[\langle M|+\langle M|\hat{U}^\dagger(\pi)\right]\left[|M\rangle+\hat{U}(\pi)|M\rangle\right]\\ &= 2\left[\langle M|M\rangle+\text{Re}\langle M|\hat{U}(\pi)|M\rangle\right] = 2\left[1+\cos(M\pi)\right],\end{aligned} \quad (B1)$$

where the unitary property of $\hat{U}$, $\cos(2\pi\mu-\pi M)=\cos(\pi M-2\pi m)=\cos(\pi M)$, and Eqs. (A3) and (A8) are used. The field amplitude of the unnormalized quantum state with $n = 4$ superposition in Eq. (1) is

$$\begin{aligned}\langle M4'|M4'\rangle &= \left\{\left[\langle M|+\langle M|\hat{U}^\dagger\left(\tfrac{\pi}{2}\right)+\langle M|\hat{U}^\dagger(\pi)+\langle M|\hat{U}^\dagger\left(\tfrac{3\pi}{2}\right)\right]\right.\\ &\quad \left.\left[|M\rangle+\hat{U}\left(\tfrac{\pi}{2}\right)|M\rangle+\hat{U}(\pi)|M\rangle+\hat{U}\left(\tfrac{3\pi}{2}\right)|M\rangle\right]\right\}\\ &= 4\left[\langle M|M\rangle+\text{Re}\langle M|\hat{U}\left(\tfrac{\pi}{2}\right)|M\rangle+\text{Re}\langle M|\hat{U}(\pi)|M\rangle+\text{Re}\langle M|\hat{U}\left(\tfrac{3\pi}{2}\right)|M\rangle\right]\\ &= 4+4\cos(\pi M)+8\cos^3\left(\tfrac{\pi}{2}M\right),\end{aligned} \quad (B2)$$

where the unitary property of $\hat{U}$, $e^{-i2\pi m}=1$, and Eqs. (A3), (A6), and (A8) are used.

## APPENDIX C: ROTATIONAL SYMMETRY OF $|Mn(\alpha)\rangle$

I describe the order of rotational symmetry for $|Mn(\alpha)\rangle$. For $n = 1$, $|M\rangle$ has the property of least rotational symmetry of one-fold $C_1$ considering its rotation by $\theta = 2\pi$ as

$$\hat{U}(2\pi)|M\rangle = |M\rangle. \quad (C1)$$

$|M\rangle$ does not have rotational symmetry with order larger than one because of the innate

character of the phase structure of a fractional optical vortex. The rotational symmetry of $|Mn(\alpha)\rangle$, $n > 1$, can be verified by the state with the superposition made up of the original and rotated states. For $n = 2$, $|M2'\rangle \equiv \hat{U}(\pi)|M\rangle + |M\rangle$ from Eq. (1). $|M2'\rangle$ satisfies the least rotational symmetry as

$$\hat{U}(2\pi)|M2'\rangle = |M2'\rangle. \tag{C2}$$

To sum $|M2'\rangle$ and its rotation state with $\theta = \pi$, I have

$$\hat{U}(\pi)|M2'\rangle + |M2'\rangle = \hat{U}(\pi)\hat{U}(\pi)|M\rangle + \hat{U}(\pi)|M\rangle + \hat{U}(\pi)|M\rangle + |M\rangle = 2\left[\hat{U}(\pi)|M\rangle + |M\rangle\right] = 2|M2'\rangle, \tag{C3}$$

where Eq. (A8) and Eq. (C1) are used. From Eq. (C3),

$$\hat{U}(\pi)|M2'\rangle = |M2'\rangle \tag{C4}$$

Owing to Eqs. (C2) and (C4), $|M2'\rangle$ has the property of rotational symmetry of $C_2$. $|M2'\rangle \equiv \hat{U}(\pi)|M\rangle + |M\rangle$ does not have the rotational symmetry with order larger than two, because $|M\rangle$ does not have rotational symmetry with order larger than one.

For $n = 4$, $|M4'\rangle \equiv |M\rangle + \hat{U}\left(\frac{\pi}{2}\right)|M\rangle + \hat{U}(\pi)|M\rangle + \hat{U}\left(\frac{3\pi}{2}\right)|M\rangle$, from Eq. (1). $|M4'\rangle$ satisfies the least rotational symmetry as

$$\hat{U}(2\pi)|M4'\rangle = |M4'\rangle. \tag{C5}$$

To sum $|M4'\rangle$ and its rotation with $\theta = \pi/2$, I have

$$|M4'\rangle + \hat{U}\left(\frac{\pi}{2}\right)|M4'\rangle = |M\rangle + \hat{U}\left(\frac{\pi}{2}\right)|M\rangle + \hat{U}(\pi)|M\rangle + \hat{U}\left(\frac{3\pi}{2}\right)|M\rangle + \hat{U}\left(\frac{\pi}{2}\right)|M\rangle + \hat{U}\left(\frac{\pi}{2}\right)\hat{U}\left(\frac{\pi}{2}\right)|M\rangle + \hat{U}\left(\frac{\pi}{2}\right)\hat{U}(\pi)|M\rangle + \hat{U}\left(\frac{\pi}{2}\right)\hat{U}\left(\frac{3\pi}{2}\right)|M\rangle = 2\left[|M\rangle + \hat{U}\left(\frac{\pi}{2}\right)|M\rangle + \hat{U}(\pi)|M\rangle + \hat{U}\left(\frac{3\pi}{2}\right)|M\rangle\right] = 2|M4'\rangle \tag{C6}$$

where Eq. (A8) and Eq. (C1) are used. From Eq. (C6),

$$\hat{U}\left(\frac{\pi}{2}\right)|M4'\rangle = |M4'\rangle. \tag{C7}$$

Similarly, to sum $|M4'\rangle$ and its rotations with $\theta = \pi$ and $\theta = 3\pi/2$, I have respectively

$$\hat{U}(\pi)|M4'\rangle = |M4'\rangle, \tag{C8}$$

and

$$\hat{U}\left(\frac{3\pi}{2}\right)|M4'\rangle = |M4'\rangle. \tag{C9}$$

Owing to Eqs. (C5), (C7)-(C9), $|M4'\rangle$ has the property of rotational symmetry of $C_4$. Similarly, $|Mn\rangle$ has the property of rotational symmetry of $C_n$ by $\hat{U}(\theta = 2\pi \times t/n)|Mn\rangle = |Mn\rangle$, $t = 1,2,...,n$. The experimental superposition can be performed using a Mach-Zehnder interferometer (MZI) with $\theta$, and it has already been verified experimentally for two light beams whose quantum states are $|M2\rangle$ and $|M4\rangle$ [14].

# APPENDIX D: USEFUL FORMULAS FOR $|\langle Mn(0)|Mn(\alpha)\rangle|^2$

$\alpha$ and $\beta \in [0, 2\pi)$.

$$\langle M(0)|\hat{U}(\beta)|M(\alpha)\rangle = \frac{e^{-im\beta}}{2\pi} \times \begin{cases} (\alpha+\beta)e^{i(2\pi-\alpha-\beta)\mu} + (2\pi-\alpha-\beta)e^{-i(\alpha+\beta)\mu}, & \alpha+\beta \leq 2\pi \\ (\alpha+\beta-2\pi)e^{i(4\pi-\alpha-\beta)\mu} + (4\pi-\alpha-\beta)e^{-i(\alpha+\beta-2\pi)\mu}, & \alpha+\beta > 2\pi \end{cases}$$

(D1)

, where Eqs. (A3) and (A5) are used. There are two expressions in two sections of Eq. (D1). Similarly,

$$\langle M(0)|\hat{U}(2\pi-\beta)|M(\alpha)\rangle = \frac{e^{im\beta}}{2\pi} \times \begin{cases} (2\pi+\alpha-\beta)e^{i(\beta-\alpha)\mu} + (\beta-\alpha)e^{-i(2\pi+\alpha-\beta)\mu}, & \alpha \leq \beta \\ (\alpha-\beta)e^{i(2\pi+\beta-\alpha)\mu} + (2\pi+\beta-\alpha)e^{-i(\alpha-\beta)\mu}, & \alpha > \beta \end{cases}.$$

(D2)

By Eqs. (D1) and (D2),

$$\langle M(\alpha)|\hat{U}(\beta)|M(0)\rangle^* = \langle M(0)|\hat{U}(2\pi-\beta)|M(\alpha)\rangle, \quad (D3)$$

where the 1-fold rotational symmetry (least rotational symmetry), unitary property of $\hat{U}$, and Eq. (A5) are used. The product of Eq. (A3) and complex conjugate of Eq. (D1) is

$$\langle M(0)|M(\alpha)\rangle\langle M(0)|\hat{U}(\beta)|M(\alpha)\rangle^*$$
$$= \frac{1}{4\pi^2} \times \begin{cases} [\alpha(\alpha+\beta)+(2\pi-\alpha)(2\pi-\alpha-\beta)]e^{i\beta M} + (2\pi-\alpha)(\alpha+\beta)e^{-i(2\pi-\beta)M} \\ +\alpha(2\pi-\alpha-\beta)e^{i(2\pi+\beta)M}, \quad \alpha+\beta \leq 2\pi \\ [\alpha(\alpha+\beta-2\pi)+(2\pi-\alpha)(4\pi-\alpha-\beta)]e^{-i(2\pi-\beta)M} + (2\pi-\alpha)(\alpha+\beta-2\pi)e^{-i(4\pi-\beta)M} \\ +\alpha(4\pi-\alpha-\beta)e^{i\beta M}, \quad \alpha+\beta > 2\pi \end{cases}.$$

(D4)

The real part of Eq. (D4) is

$$\text{Re}\left[\langle M(0)|M(\alpha)\rangle\langle M(0)|\hat{U}(\beta)|M(\alpha)\rangle^*\right]$$

$$= \frac{1}{4\pi^2} \times \begin{cases} \left[4(\alpha^2 + \alpha\beta - 2\pi\alpha - \pi\beta)\sin^2(\pi M) + 4\pi^2\right]\cos(\beta M) + 2\pi\beta\sin(2\pi M)\sin(\beta M),\ \alpha+\beta \leq 2\pi \\ 2(\alpha^2 + \alpha\beta - 4\pi\alpha - \pi\beta + 4\pi^2)\cos[(2\pi-\beta)M] + (2\pi-\alpha)(\alpha+\beta-2\pi)\cos[(4\pi-\beta)M] \\ +\alpha(4\pi-\alpha-\beta)\cos(\beta M),\ \alpha+\beta > 2\pi \end{cases}$$

(D5)

By Eq. (D1),

$$\langle M(0)|\hat{U}(\beta_1)|M(\alpha)\rangle\langle M(0)|\hat{U}(\beta_2)|M(\alpha)\rangle^*$$

$$= \frac{e^{-i(\beta_1-\beta_2)M}}{4\pi^2} \times \begin{cases} 4\sin^2(\pi M)\left[(\alpha+\beta_1)(\alpha+\beta_2) - \pi(2\alpha+\beta_1+\beta_2)\right] + 4\pi^2 \\ +i2\pi(\beta_1-\beta_2)\sin(2\pi M),\ \alpha+\beta_1 \leq 2\pi\ \text{and}\ \alpha+\beta_2 \leq 2\pi \\ \left[(\alpha+\beta_1)(\alpha+\beta_2-2\pi) + (2\pi-\alpha-\beta_1)(4\pi-\alpha-\beta_2)\right]e^{-i2\pi M} \\ +(2\pi-\alpha-\beta_1)(\alpha+\beta_2-2\pi)e^{-i4\pi M} + (\alpha+\beta_1)(4\pi-\alpha-\beta_2) \\ ,\ \alpha+\beta_1 \leq 2\pi\ \text{and}\ \alpha+\beta_2 > 2\pi \\ \left[(\alpha+\beta_1-2\pi)(\alpha+\beta_2) + (4\pi-\alpha-\beta_1)(2\pi-\alpha-\beta_2)\right]e^{i2\pi M} \\ +(\alpha+\beta_1-2\pi)(2\pi-\alpha-\beta_2)e^{i4\pi M} + (4\pi-\alpha-\beta_1)(\alpha+\beta_2) \\ ,\ \alpha+\beta_1 > 2\pi\ \text{and}\ \alpha+\beta_2 \leq 2\pi \\ 4\sin^2(\pi M)\left[(\alpha+\beta_1)(\alpha+\beta_2) - \pi(6\alpha+3\beta_1+3\beta_2-8\pi)\right] + 4\pi^2 \\ +i2\pi(\beta_1-\beta_2)\sin(2\pi M),\ \alpha+\beta_1 > 2\pi\ \text{and}\ \alpha+\beta_2 > 2\pi \end{cases}$$

(D6)

There are four expressions in four sections of Eq. (D6). As $\beta_1 = \beta_2 = \beta$, Eq. (D6) becomes

$$\left|\langle M(0)|\hat{U}(\beta)|M(\alpha)\rangle\right|^2 = \frac{1}{4\pi^2} \times \begin{cases} 4(\alpha+\beta)(\alpha+\beta-2\pi)\sin^2(\pi M) + 4\pi^2,\ \alpha+\beta \leq 2\pi \\ 4(\alpha+\beta-4\pi)(\alpha+\beta-2\pi)\sin^2(\pi M) + 4\pi^2,\ \alpha+\beta > 2\pi \end{cases}.$$

(D7)

The real part of Eq. (D6) is

$$\mathrm{Re}\left[\langle M(0)|\hat{U}(\beta_1)|M(\alpha)\rangle\langle M(0)|\hat{U}(\beta_2)|M(\alpha)\rangle^*\right]$$

$$=\frac{1}{4\pi^2}\times\begin{cases}\{4\sin^2(\pi M)[(\alpha+\beta_1)(\alpha+\beta_2)-\pi(2\alpha+\beta_1+\beta_2)]+4\pi^2\}\cos[(\beta_1-\beta_2)M]\\+2\pi(\beta_1-\beta_2)\sin(2\pi M)\sin[(\beta_1-\beta_2)M],\ \alpha+\beta_1\leq 2\pi\text{ and }\alpha+\beta_2\leq 2\pi\\[(\alpha+\beta_1)(\alpha+\beta_2-2\pi)+(2\pi-\alpha-\beta_1)(4\pi-\alpha-\beta_2)]\cos[(\beta_1-\beta_2+2\pi)M]\\+(2\pi-\alpha-\beta_1)(\alpha+\beta_2-2\pi)\cos[(\beta_1-\beta_2+4\pi)M]+(\alpha+\beta_1)(4\pi-\alpha-\beta_2)\cos[(\beta_1-\beta_2)M]\\,\ \alpha+\beta_1\leq 2\pi\text{ and }\alpha+\beta_2>2\pi\\[(\alpha+\beta_1-2\pi)(\alpha+\beta_2)+(4\pi-\alpha-\beta_1)(2\pi-\alpha-\beta_2)]\cos[(\beta_1-\beta_2-2\pi)M]\\+(\alpha+\beta_1-2\pi)(2\pi-\alpha-\beta_2)\cos[(\beta_1-\beta_2-4\pi)M]+(4\pi-\alpha-\beta_1)(\alpha+\beta_2)\cos[(\beta_1-\beta_2)M]\\,\ \alpha+\beta_1>2\pi\text{ and }\alpha+\beta_2\leq 2\pi\\\{4\sin^2(\pi M)[(\alpha+\beta_1)(\alpha+\beta_2)-\pi(6\alpha+3\beta_1+3\beta_2-8\pi)]+4\pi^2\}\cos[(\beta_1-\beta_2)M]\\+2\pi(\beta_1-\beta_2)\sin(2\pi M)\sin[(\beta_1-\beta_2)M],\ \alpha+\beta_1>2\pi\text{ and }\alpha+\beta_2>2\pi\end{cases}.$$

(D8)